# Impact of Recent Discoveries on Petroleum and Natural Gas Exploration: Emphasis on India

by

**J. Marvin Herndon**
**Transdyne Corporation**
**San Diego, CA 92131 USA**

mherndon@san.rr.com

**Abstract:** Two recent discoveries greatly impact understanding relevant to the origination and emplacement of petroleum and natural gas deposits. One discovery, pertaining to hydrocarbon formation from methane, broadens significantly potential regions where abiotic petroleum and natural gas deposits might be found. The other, discovery of the physical impossibility of Earth-mantle convection, restricts the range and domain of geodynamic behavior, and leads to new insights on the formation of petroleum and natural gas deposits. This article highlights the impact and implications of those discoveries, especially as they relate to petroleum and natural gas exploration in India and throughout the world. From the reasoning developed here, the generality of the considerations involved, the understanding developed with respect to the East African Rift System, and the experience garnered from the larger and older Siberian Traps, the prognosis and potential for the region beneath the Deccan Traps of India to eventually become a major source of petroleum and natural gas seems to the author to be quite favorable.

## Introduction

*In situ* geophysical measurements and geological understanding are the keys to successful petroleum and natural gas exploration. But, for decades understanding has been hampered by potential unknowns related to the origin of petroleum and natural gas, and by certain misconceptions related to the geodynamic behavior of the Earth. Two recent discoveries, though, impact in fundamental and positive ways geological understanding relevant to petroleum and natural gas exploration and eventual exploitation. One, the experimental verification of hydrocarbon formation from methane, $CH_4$, at upper-mantle temperatures and pressures [1], broadens significantly potential regions where abiotic petroleum and natural gas deposits might be found. The other, Herndon's discovery of the physical impossibility of mantle convection [2], restricts considerably the range and domain of geodynamic behavior relevant to the origination and emplacement of petroleum and natural gas deposits. The purpose of this article is to review the impact and implications of those two scientific discoveries, especially as they relate to petroleum and natural gas exploration in India and throughout the world.

## Origin of Petroleum and Natural Gas

The idea that a significant proportion of petroleum and natural gas resources are of non-biological origin is still considered controversial, but increasingly there is discussion, observation, and experimentation. The seminal concept of abiotic petroleum and natural gas being primordial matter erupted from deep within the Earth was announced at the beginning of the 19th Century by Alexander von Humboldt and Louis Joseph Gay-Lussac. In 1951, Kudryavtsev [3] originated the modern Russian-Ukrainian theory of abiotic petroleum and natural gas. Laboratory experiments demonstrating the feasibility of hydrocarbon formation under deep-Earth conditions, even in the absence of primary hydrocarbons, provided the first major support for that theory [4], and observational evidence is beginning to be reported [5, 6].

As noted by Herndon [7], ultimately the prognosis for vast potential resources of mantle and deep-crust petroleum and natural gas depends critically upon the nature and circumstances of Earth's formation. In that General Article in *Current Science*, Herndon showed that the popular "standard model of solar system formation" would lead to the contradiction of terrestrial planets having insufficiently massive cores. He presented evidence in support of the idea that the Earth initially formed as a gas-giant planet by raining out from within a giant gaseous protoplanet. Prior to being stripped of its gas envelope, the proto-Earth was nearly 300 times as massive as present-day Earth, almost identical to Jupiter. Notably, the evidence points to the entirety of Earth formation having taken place in intimate association with primordial gases, which includes about 1.3 Earth-masses of methane, $CH_4$. The possibility of carbon-compound occlusion under these conditions is greatly enhanced, relative to the previous Earth-formation concept, and is decidedly relevant to the enhanced prognosis for a deep-Earth methane reservoir and for a carbon-source for abiotic petroleum and natural gas.

Previously, experiments aimed at demonstrating the formation of *n*-alkane hydrocarbons under at-depth thermodynamic conditions, ca. 1,500 °C at 50 kbar, involved the use of calcium carbonate, $CaCO_3$, for source-carbon [4]. Recently, though, Kolesnikov et al. [1] showed that, when methane is exposed to pressures higher than 2 GPa and to temperatures in the range of 700-1200 °C, it reacts to form hydrocarbons containing 2-4 carbon atoms. The results of that investigation not only add experimental support to the concept of abiotic petroleum formation, but advance the possibility of its formation directly from methane, rather than from $CaCO_3$.

When plants photosynthetically utilize atmospheric carbon dioxide, $CO_2$, isotope fraction occurs. Plants preferentially retain the lighter carbon isotope, $^{12}C$. The observed variation range of $\delta^{13}C$ in petroleum, being similar to the range of $\delta^{13}C$ variation in plant carbon, is sometimes used as a basis to assume a biogenic origin for petroleum. Such similarity, though, cannot be taken as proof that petroleum exclusively is of biogenic origin. As discussed below, the circumstances for trapping mantle-derived, abiotic petroleum are the same as for biogenic petroleum, consequently, sample contamination would be inevitable. Moreover, the isotopic composition of deep-Earth methane is not known, and would certainly change during transit through porous media, becoming progressively lighter by diffusion fractionation [8]. Circumstantially, the non-carbonate carbon of carbonaceous meteorites is isotopically lighter than atmospheric carbon [9], as are deep-mantle diamonds [10]. At present, there is no known way from measurements of stable carbon isotope ratios to characterize unambiguously the origin of petroleum.

The experimental formation of *n*-alkane hydrocarbons from methane under Earth-mantle conditions by Kolesnikov et al. [1] did not address whether or in what manner kinetic isotope fractionation effects might occur. Hopefully, that will be done in future experiments as petroleum formation from methane is entirely consistent with the enhanced prognosis for a deep-Earth methane reservoir described previously [7] and becomes increasingly significant in light of the evidence discussed below.

## Physical Impossibility of Mantle Convection

During World War II, Hess [11] allowed his ship's new echo-sounding equipment to run continuously during long transits across the Pacific Ocean, providing for the first time, lengthy profiles of the ocean floor. He correctly guessed that as new seafloor forms at the mid-oceanic ridges, the older seafloor slides across the expanse of ocean, eventually plunging into submarine trenches. Hess believed that this was just the visible part of a great conveyer-belt loop caused by convection in the Earth's mantle [11]. Mantle convection thus became a crucial component of his seafloor spreading theory and, subsequently, a crucial component of plate tectonics theory, which wholly adopted seafloor spreading.

Elsasser [12] had paved the way for ready acceptance of deep-Earth convection by having set forth in 1939 his idea that the geomagnetic field is generated by convection-driven dynamo action in the Earth's fluid core. That the mantle is solid, not liquid, was not a serious concern because over long periods of time, for example, glass church-windows are observed to sag and the mantle was thought to be somewhat plastic. Moreover, calculation of the dimensionless Rayleigh Number [13] using realistic mantle parameters invariably yielded a high value supposedly indicative of convection. Furthermore, plate tectonics theory seemed to explain the ocean floor observations so well that many thought mantle convection "must exist". For decades, the idea of mantle convection has underlain virtually all geophysical considerations involving the Earth's mantle. But there is a very serious problem discovered by Herndon [2]: Earth-mantle convection, like Earth-core convection [14], is physically impossible.

Chandrasekhar [15] described convection in the following way: "The simplest example of thermally induced convection arises when a horizontal layer of fluid is heated from below and an adverse temperature gradient is maintained. The adjective 'adverse' is used to qualify the prevailing temperature gradient, since, on account of thermal expansion, the fluid at the bottom becomes lighter than the fluid at the top; and this is a top-heavy arrangement which is potentially unstable. Under these circumstances the fluid will try to redistribute itself to redress this weakness in its arrangement. This is how thermal convection originates: It represents the efforts of the fluid to restore to itself some degree of stability." The clarity of Chandrasekhar's explanation inspired Herndon [2] to think-upon and, later, to discover why convection in the Earth's mantle is physically impossible.

Frequent reference is (wrongly) made to the parameters of the Earth's mantle yielding a high Rayleigh Number, supposedly indicative of vigorous convection. In 1916, Lord Rayleigh [13] applied the Boussinesq [16] approximation to Eulerian equations of motion to derive that dimensionless number to quantify the onset of instability in a thin, horizontal layer of fluid heated from beneath. The underlying assumptions, however, are inconsistent with the physical parameters of the Earth's mantle, viz.; Earth's mantle being "incompressible", density being "constant" except as modified by thermal expansion, and pressure being "unimportant" (quotes from Lord Rayleigh [13]).

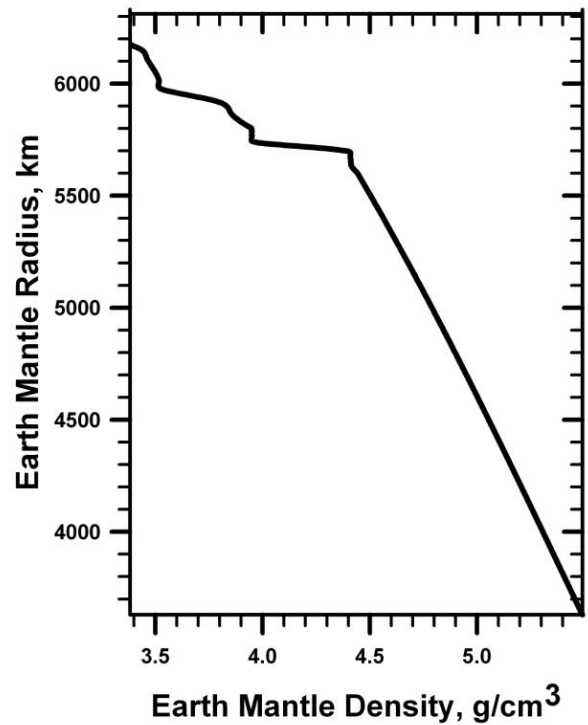

**Figure 1.** Density as a function of radius in the Earth's mantle [17, 18].

As shown in Figure 1, because of its own weight and the weight of the crust, the Earth's mantle is about 62% denser at the bottom than at the top [17, 18]. The tiny amount of thermal expansion that may occur at the mantle-bottom, < 1%, cannot overcome such a great difference in density, meaning the Earth's mantle cannot become top-heavy and meaning also that a thermally expanded "parcel" of bottom-mantle matter cannot become light enough to float to the mantle-top. The Earth's mantle, therefore, cannot convect.

In the natural physical sciences, it is difficult, some might say impossible, to prove a proposition to be true; but, on the other hand, it is quite possible to prove that which is false. The author has proven the proposition of mantle convection to be false [2]. Thus, all geophysical considerations which are based upon the assumption of mantle convection need to be revised.

## Geology without Mantle Convection

Modern-day geological interpretations are frequently guided and/or directed by inferences from plate tectonics. But now, the author can state unequivocally that plate tectonics theory is incorrect [2]. After its having been the dominant, widely accepted geodynamic theory for more than forty years, plate tectonics theory is incorrect because it depends critically upon the assumption of mantle convection which is physically impossible. In certain respects, the failure of plate tectonics should not be too surprising. Plate tectonics theory not only was based upon unobserved, assumed convection, but was formulated without knowledge of an energy source sufficient to power geodynamic activity throughout the time Earth has existed.

The seemingly compelling agreement between seafloor observations and plate tectonics arises as a consequence of the global dynamics described by Herndon's *whole-Earth decompression dynamics* [19, 20], which does not require or depend upon mantle convection. Briefly, whole-Earth decompression dynamics is the consequence of Earth having formed as a Jupiter-like gas giant. Beneath about 300 Earth-masses of hydrogen, helium, and volatile compounds, such as methane, the silicate-rock-plus-alloy kernel was compressed to about 64% of present-day radius, an amount sufficient to have yielded a closed, contiguous shell of continental rock without ocean basins.

After being stripped of its great overburden of volatile protoplanetary gases by the high temperatures and/or by the violent activity of the T Tauri-phase solar wind, associated with the thermonuclear ignition of the Sun, the Earth would inevitably begin to decompress, to rebound toward a new hydrostatic equilibrium. The initial whole-Earth decompression is expected to result in a global system of major *primary* cracks appearing in the rigid crust which persist and are identified as the global, mid-oceanic ridge system, just as explained by Earth expansion theory [21, 22]. But here the similarity with that theory ends. Whole-Earth decompression dynamics [19], sets forth a different mechanism for global geodynamics which involves the formation of *secondary* decompression cracks and the in-filling of those cracks, a process which is not limited to the last 180 million years, the maximum age of the seafloor.

As the Earth subsequently decompresses and swells from within, the deep interior shells may be expected to adjust to changes in radius and curvature by plastic deformation. As the Earth decompresses, the area of the Earth's rigid surface increases by the formation of secondary decompression cracks often located near the continental margins and presently identified as submarine trenches. These secondary decompression cracks are subsequently in-filled with basalt, extruded from the mid-oceanic ridges, which traverses the ocean floor by gravitational creep, ultimately plunging into and in-filling secondary decompression cracks, thus emulating the plate-tectonic-concept of subduction, but without mantle convection and without re-circulation of oceanic crust through the mantle as previously assumed in plate tectonics theory.

The ancient supercontinent, Pangaea, envisioned by Wagener [23] and adopted in plate tectonic considerations, was thought to be surrounded by ocean occupying nearly 1½ times its surface area. By contrast, in whole-Earth decompression dynamics, there is but one true ancient supercontinent, the 100% closed contiguous shell of continental rock called Ottland in honor of Ott Christoph Hilgenberg, who first conceived of its existence [22]. Moreover, there can never be another. The fragmentation of Ottland by primary decompression cracks and the formation of interstitial ocean basins in conjunction with secondary decompression cracks need not take place immediately upon removal of primordial gases, nor necessarily occur at a single point in time. The yet-to-be-delineated sequences of Ottland subdivision and displacement may bear only superficial resemblance to the popular, but hypothetical, breakup of Pangaea, where the continents are assumed free to wander, breaking up and re-aggregating, while riding atop non-existent mantle convection cells. In whole-Earth decompression dynamics there are fewer degrees of freedom. While complex, there is simplicity as well because continental fragmentation and dispersion represents the approach toward dynamical equilibrium, not the random breaking up and re-aggregating previously imagined.

Much of geology and geophysics is built upon the assumption of mantle convection and the making of models based upon that assumption and the making of models based upon those models. The physical impossibility of mantle convection means, for example, that there is no reason to believe that the mantle has been repeatedly turned over and mixed, and there is no reason to believe that the mantle has been significantly degassed. The latter circumstance greatly enhances the prognosis for future abiotic energy supplies, at least by comparison to previous ideas [7].

Seismological investigations have revealed the internal physical structure of Earth [24-26]. More recently, with the establishment of extensive seismological networks and powerful supercomputing technology, seismic tomography has begun to image structures within the mantle [27-29]. Because the raw data is limited, consisting of P-wave and S-wave velocities, the images produced are subject to interpretations, often based upon pre-perceived Earth behavior. For example, one finds image descriptions, colored by belief in mantle convection, such as "down-plunging slabs", which can be interpreted instead as "in-filled secondary decompression cracks".

Tomographic images of so-called mantle plumes have become increasingly important for geological understanding. But, even with the advent of seismic tomography, there is still considerable controversy as to the true nature of mantle plumes and to the question of whether or not mantle plumes actually exist [30, 31].

The mantle plume concept had its origins in Wilson's 1963 suggestion [32] that the volcanic arc comprised of the Hawaiian Islands formed as seafloor moved across a persistent, fixed "hot-spot". In 1971, Morgan [33] proposed that hot-spots are manifestations of convection in the lower mantle.

In certain instances, hot-spots extending to the base of the lower mantle have been imaged by seismic tomography, further encouraging the obsessive idea of mantle convection [27, 29]. After decades of hot-spot controversy, though, the author can now state unequivocally that mantle plumes are not manifestations of mantle-convection, because convection in the Earth's mantle is physically impossible [2]. Mantle plumes involve thermal transport, not mass transport. Thus, the range and domain of geological understanding, particularly relevant to petroleum and natural gas exploration, is once again further restricted.

The decades-long controversy surrounding hot-spots and mantle plumes has had the consequence of generating a plethora of confusing, contradictory geological interpretations, often based upon the invalid proposition of mantle convection and/or upon plate-tectonic, model-dependent assumptions. Set those aside and consider anew the behavior of Earth, particularly, its surface manifestations.

## Herndon's New Earth Dynamics

Globally, virtually all major geological activity is the consequence of a single process, the formation of new surface area to accommodate decompression-increased planetary volume, which primarily involves crustal extension fracturing, basalt extrusion, and decompression crack in-filling [19, 20]. The principal consequence is the splitting of Earth's surface by two types of decompression cracks. Primary decompression cracks are underlain by a heat-source(s) sufficient to generate and to extrude molten basalt. Secondary decompression cracks, on the other hand, lacking such a heat-source(s), produce trenches into which basalt, extruded elsewhere, ultimately in-fills. Energy for geological activity is primarily the vast, stored energy of protoplanetary compression, augmented to some extent by georeactor nuclear fission energy [34-37] and by radioactive decay energy.

Mantle decompression will tend to propagate throughout the mantle, like a tsunami, until it reaches the impediment posed by the base of the crust. There, crustal rigidity opposes continued decompression; pressure builds and compresses matter at the mantle-crust-interface, resulting in compression heating. Ultimately, pressure is released at the surface through primary decompression crack formation, *i.e.*, extension fracturing with its associated basin formation and volcanism, and through secondary decompression crack formation and/or enlargement. The process of mantle decompression thermal-tsunami may account for much of the heat leaving the Earth's surface, for the geothermal gradient observed in the crust, for substantial volcanism and, perhaps, for earthquake generation as well [20]. That process also might greatly enhance the prognosis for future abiotic energy supplies by pressurizing and heating the base of the crust, a potential collection point for mantle methane or other carbonaceous matter [7]. One consequence of Herndon's discovery of the physical impossible of mantle convection [2] is the realization that the whole-Earth decompression and the phenomena previously called "mantle plumes" are intimately and actively connected.

In the 1960's geoscientists discovered occluded helium in oceanic basalts which, remarkably, possessed a higher $^3$He/$^4$He ratio than air. At the time there was no known deep-Earth mechanism that could account for the experimentally measured $^3$He, so its origin was assumed to be a primordial $^3$He component, trapped at the time of Earth's formation, which was subsequently diluted with $^4$He from radioactive decay. State-of-the-art numerical simulations of georeactor operation, conducted at Oak Ridge National Laboratory, yielded fission-product helium, as shown in Figure 2, with isotopic compositions within the exact range of compositions typically observed in oceanic basalts [36, 38]. For additional information, see Rao [39].

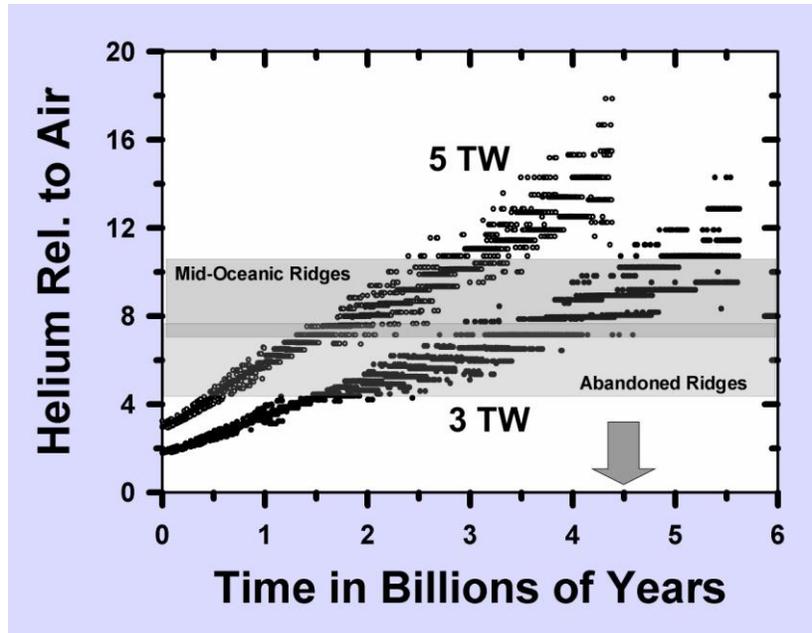

**Figure 2.** Fission product ratio of $^3$He/$^4$He, relative to that of air, $R_A$, from nuclear georeactor numerical calculations at 5 TW (upper) and 3 TW (lower) power levels [36]. The band comprising the 95% confidence level for measured values from mid-oceanic ridge basalts (MORB) is indicated by the solid lines. The age of the Earth is marked by the arrow. Note the distribution of calculated values at 4.5 Gyr, the approximate age of the Earth. The increasing values are the consequence of uranium fuel burn-up. Iceland deep-source "plume" basalts present values ranging as high as 37 $R_A$ [40]. Figure from [41].

Forty-nine hot-spots have been identified, but not all have identical surface or compositional manifestations [42]. About 18% of those, including the Hawaiian Islands and Iceland, are characterized by basalts containing helium with high $^3$He/$^4$He ratios, *i.e.*, $R_A$>10. Beneath both Hawaii and Iceland, seismic tomography images plume-like structures which extend downward to the base of the lower mantle [27-29]. Because mantle convection is physically impossible [2], these structures cannot be plumes, not matter transported by buoyancy from the bottom of the lower mantle to the surface. Instead, these appear to be thermal in origin, paths where heat from the Earth's core channels to the surface.

Helium, being a very light, highly mobile, inert gas, apparently can travel from Earth's core to surface through these heat conduits, progressing upward toward ever-decreasing densities. In addition to portending the eventual demise of the geomagnetic field [36], the high $^3$He/$^4$He ratios observed serve as markers, identifying the particular basalt as one whose extrusion involved thusly channeled Earth-core heat.

Rather than basalt extrusion occurring simply as seafloor passes over a hot-spot, global-scale, whole-Earth decompression appears to be intrinsically involved, at least in instances where Earth-core channeled heat can be identified by high $^3$He/$^4$He ratios. Mjelde and Faleide [43] have discovered, for example, variations in Icelandic basalt production which have the same periodicity and relative timing as Hawaiian basalt variations on the opposite side of the Earth. Their discovery may have broad, but yet unappreciated, implications.

Seafloor passing over hot-spots offers a reasonable explanation for the formation of basaltic island arcs, such as the Hawaiian Islands and the associated Emperor Seamount chain. Significantly, that explanation is in accord with whole-Earth decompression dynamics. The generalization of the "passing over a hot-spot" concept to mid-continental basalt floods, such as the Deccan Traps in India and the Siberian Traps in central Russia, both of which have high $^3$He/$^4$He ratios [44, 45], though, is untenable in light of impossible plate tectonics [2].

## Implications

For four decades, petroleum exploration geology has been described in terms of plate tectonics, which is based upon physically-impossible mantle convection [2]. Surface manifestations, such as foreland basins and back-arc basins, which seem well-described by plate tectonics, are in many instances similarly described by whole-Earth decompression dynamics, although with subtle, but important, differences, especially those related to the absence of subduction.

In plate tectonics terminology, "rift" refers to the interface of two plates that are beginning to pull apart. In whole-Earth decompression dynamics, "rift" refers to the beginning of the formation of a decompression crack. Rifting is an integral part of the whole-Earth decompression process of successive continent fragmenting, beginning with Ottland and continuing into the present, and about which the author makes the following generalizations.

The process of continent fragmenting begins with the formation of a decompression crack. Over time, the crack widens, forming a rift-valley or basin. Volcanic eruptions may subsequently occur, depending mainly upon available heat. The rift-basin thus formed becomes an ideal environment for the development of geological strata frequently associated with petroleum and natural gas deposits and can remain a part of the continental margins even after ocean floor formation.

Virtually all petroleum deposits are connected in some way to, or are the consequence of, rifting, even those deposits, such as foreland basins, that involve underthrust compression, which may result from rifting and extension elsewhere. Continent fragmenting, both successful and failed, initiates with rifting. Observations of rifting which is currently taking place at the Afar triangle in northeastern Ethiopia, and observations of the consequences of rifting throughout the East African Rift System (Figure 3) can help to shed light on the nature of petroleum-deposit related rifting that has occurred elsewhere.

The Afar triangle is the triple junction where the Red Sea rift, the Carlsberg Ridge of the Indian Ocean, and the East African Rift System meet. Seismic tomographic imaging beneath that region shows a very large, low-velocity zone extending to the base of the lower mantle, referred to as a "superplume" [46, 47]. Because heat at the base of the lower mantle cannot make bottom-mantle matter sufficiently buoyant to float upward [2], the high $^3$He/$^4$He ratios, $R_A>10$, measured in Afar volcanic basalt [48], indicate mantle-deep heat channeling, which allows the highly mobile, inert helium to migrate upward toward regions of progressively-lower density.

The extension-related processes observed at Afar and along the East African Rift System provide all of the crucial components for petroleum-deposit formation. Rifting causes the formation of deep basins, as evidenced, for example, by Lake Tanganyika, depth 1.4 km, the second deepest lake in the world, and by Lake Nyasa, depth 0.7 km, the fourth deepest lake, both of which occur as part of the East African Rift System. The observed uplifting caused by swelling from below [49] makes surface land susceptible to erosion, thus providing great amounts of sedimentary material for reservoir rock in-filling of basins. Volcanic-derived sedimentary material may be richly mineralized, assuring strong phytoplankton blooms. Uplifting can sequester sea-flooded lands thus leading to the formation of marine halite deposits through desiccation, and can also lead to dome formation as well.

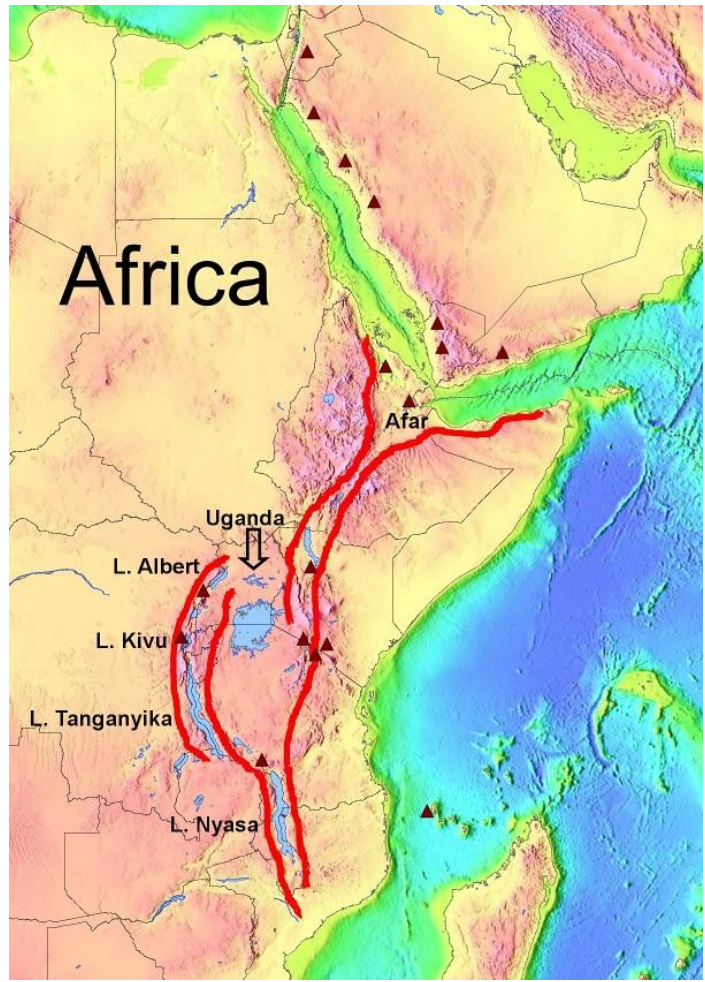

**Figure 3.** Northeastern portion of Africa. Red lines show the major rifts comprising the East African Rift System. Active volcanoes are indicated by maroon triangles.

Afar swelling is important in providing sediments for reservoir sands, delivered by the Congo, Niger and Nile rivers, to continent-edge basins, which were formed during earlier continent fragmenting by extension and which are Africa's main petroleum provinces. All of the components of reservoir, source and seal are related to the influx of sediments [50]. Local rifting can provide all those components, as well as basin formation, as indicated by a report published in June 2, 2009 by *East African Business Week* which notes that, based upon recent test-well results, a senior official of the U. S. Department of Energy indicated that Uganda's [Lake Albert basin of the East African Rift System] oil reserves could be as much as that of Persian Gulf countries.

And, what of the possibility of abiotic petroleum? Although there is not yet an unambiguous way to ascertain the extent to which petroleum might be of abiotic origin, the possibility, though controversial, should not be dismissed, especially in circumstances where rifting is involved. For example, in 1980 Gold and Soter [51] stated that Lake Kivu, part of the East African Rift System, contains 50 million tons of dissolved methane for which there is no adequate microbial source.

The region, including the West Siberian Basin, located between the Ural Mountains and the Siberian Platform, and the Siberian Platform, underwent extensive rifting about 500-250 million years ago and uplifting [52], leading to rift-basin formations that developed geological strata on a grand scale, extremely conducive for trapping petroleum and natural gas [53]. About 250 million years ago, massive basalt floods spewed forth for about one million years, blanketing the area with perhaps more than 2,000 km$^3$ of basalt containing helium having high $^3$He/$^4$He ratios [44]. These basalts are known as the Siberian Traps. Evidence indicates that rifting continued after basalt extrusion [53]. Rifting appears to underlie the formation of rift-basins and later the massive eruption of basalt. The high $^3$He/$^4$He ratios indicate that heat was channeled from the Earth's core. Today, that area is known to contain some of the most extensive petroleum, natural gas, and coal deposits in the world.

India can be understood geologically as a continent in the process of fragmenting, the specifics of which are described in detail by Sheth [54]. Figure 4, the schematic representation of India, adapted from Sheth [54], emphasizes her numerous major rifts. The author has added to that figure in magenta the rift-related locations where major petroleum and natural gas discoveries have been made. Rift-basin formations tend to develop geological strata conducive for trapping petroleum and natural gas.

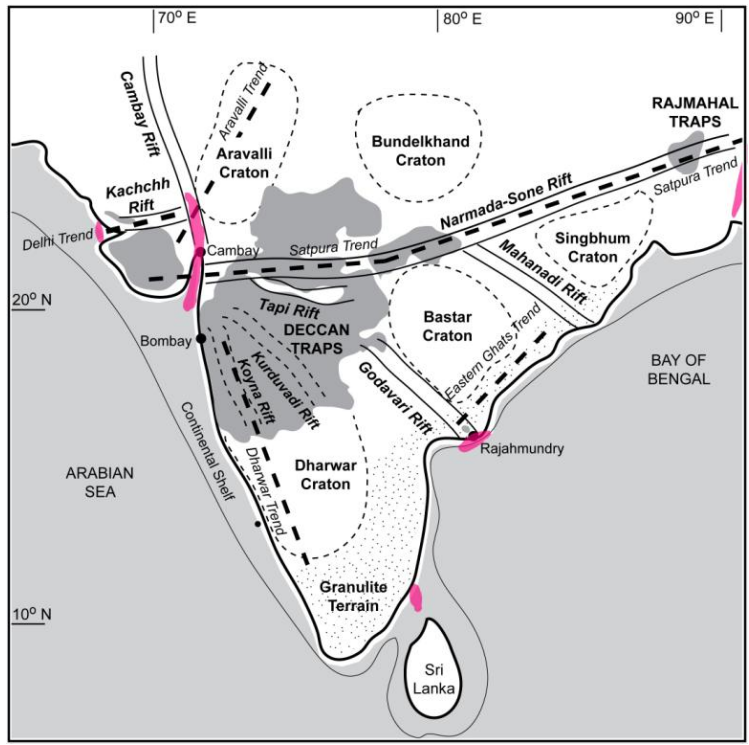

**Figure 4.** Schematic representation of India, adapted with permission from Sheth [54] showing rift zones crossing peninsula India and showing the present outcrop areas of the Deccan and Rajmahal flood basalts (shaded). Areas where significant rift-related petroleum and natural gas discoveries have been made are shown in magenta. Not to scale.

Sheth [54] set forth compelling evidence that India's Deccan Traps originated by rifting. The consequence of rifting is the formation of rift-valleys and rift-basins, which over geological time may develop into sedimentary basins with geological strata favorable to the entrapment of petroleum and natural gas. Like the Siberian Traps, the Deccan Traps are the consequence of massive flood basalt eruptions about 65 million years ago which blanketed the underlying geological features. The high $^3$He/$^4$He ratios observed in Deccan-basalt is indicative of the heat having been channeled from the Earth's core [45], which is consistent with the observation of a deep seismic low-velocity zone beneath [55].

An article in the May 1, 2008 issue of *The Economic Times* announced the discovery by ONGC of petroleum and natural gas beneath the Deccan Traps. From the standpoint of the reasoning developed in this paper, the generality of the considerations involved, the understanding developed with respect to the East African Rift System, and the experience garnered from the larger and older Siberian Traps, the prognosis and potential for the Deccan Traps to eventually become a major source of petroleum and natural gas seems to the author to be quite favorable.

Generally, continent fragmenting, as the result of whole-Earth decompression and channeled heat from beneath, whether on-going, completed, or arrested, leads to rift-basin formation, to surface uplift and doming, and to the formation of voluminous volcanic deposits. The geological consequences of which, as described in this article, can provide all of the crucial components for petroleum and natural gas deposits: basin, reservoir, source and seal. Not surprisingly, much oil and gas exploration activities are focused along continental margins where fragmenting and, presumably petroleum deposit formation, has taken place. But what of instances when continent masses, fragmented by rifting, are forced back together by competitive interactions caused by extensions elsewhere? It is tempting to speculate as to whether the foreland petroleum and natural gas deposits of the Assam and Tripura regions of northeastern India might have begun by rifting during continent fragmentation before India was forced into collision with Asia and formed foreland basins.